\documentclass[a4paper,12pt]{article}
\usepackage{amsmath,amssymb,bm,cite,graphicx,subfigure}
\usepackage[setpagesize=false]{hyperref}

\renewcommand{\arraystretch}{1.2}

\makeatletter
 
 \@addtoreset{equation}{section}
\makeatother

\begin{document}
\thispagestyle{empty}


\leftline{\today \hfill OU-HET-1107}
\rightline{KYUSHU-HET-229}

\vskip3.0cm

\baselineskip=30pt plus 1pt minus 1pt
\begin{center}
{\LARGE \bf Signals of $W'$ and $Z'$ bosons at the LHC \\ in the $SU(3)\times SO(5)\times U(1)$ gauge-Higgs unification}
\end{center}

\baselineskip=22pt plus 1pt minus 1pt
\vskip 2.0cm

\begin{center}
{\bf Shuichiro Funatsu$^1$, Hisaki Hatanaka$^2$, Yutaka Hosotani$^3$,}

{\bf Yuta Orikasa$^4$ and Naoki Yamatsu$^5$}

\baselineskip=17pt plus 1pt minus 1pt
\vskip 10pt
{\small \it $^1$Institute of Particle Physics and Key Laboratory of Quark and Lepton Physics (MOE), Central China Normal University, Wuhan, Hubei 430079, China} \\
{\small \it $^2$Osaka, Osaka 536-0014, Japan} \\
{\small \it $^3$Department of Physics, Osaka University, Toyonaka, Osaka 560-0043, Japan} \\
{\small \it $^4$Institute of Experimental and Applied Physics, Czech Technical University in Prague,} \\
{\small \it Husova 240/5, 110 00 Prague 1, Czech Republic} \\
{\small \it $^5$Department of Physics, Kyushu University, Fukuoka 819-0395, Japan} \\
\end{center}

\vskip 2.0cm

\begin{abstract}
    The $pp\to \{W, W'\} \to l\nu$ and $pp\to \{\gamma, Z, Z'\} \to l^+l^-$ ($l=e,\mu$) processes in the $SU(3)_C\times SO(5)\times U(1)_X$ gauge-Higgs unification (GHU) models are studied, 
    where $W'$ and $Z'$ bosons are Kaluza-Klein (KK) exited states of the electroweak gauge bosons. 
    From the experimental data collected at the Large Hadron Collider, constraints on the KK mass scale and the Aharonov-Bohm phase are obtained. 
    One can explore the KK mass scale in the GUT inspired GHU model up to 18 TeV for the luminosity 300 fb$^{-1}$ and 22 TeV for the luminosity 3000 fb$^{-1}$ at $\sqrt{s}=14$ TeV. 
\end{abstract}

\newpage

\baselineskip=20pt plus 1pt minus 1pt
\parskip=0pt

\section{Introduction}
Experimental results at the Large Hadron Collider (LHC) collected at a center-of-mass energy $\sqrt{s} = 13$ TeV during the year 2015-2018 
have been presented by ATLAS~\cite{ATLAS:2019erb,ATLAS:2020yat,ATLAS:2019lsy,ATLAS:2019fgd} and CMS~\cite{CMS:2021ctt,CMS:2019gwf} groups. 
No direct signals of new physics beyond the standard model (SM) have been observed at the LHC so far. 
In many models such as the sequential standard model, left-right symmetric model, grand unified theories (GUT) and models with an extra dimension, there appear 
$W'$ or $Z'$ bosons~\cite{Leike:1998wr,Langacker:2008yv,Csaki:2015hcd,Zyla:2020zbs}. 
Physics of $W'$ and $Z'$ bosons at the LHC has been an important subject~\cite{Accomando:2011eu,Accomando:2013sfa,Accomando:2019ahs,
Agashe:2016kfr,Barducci:2012kk,Greco:2014aza,Liu:2018hum,Edelhauser:2013lia,Deutschmann:2017bth,Han:2012vk,Chiang:2014yva,Pappadopulo:2014qza,Fairbairn:2016iuf,Das:2016zue,Mitra:2016kov,Amrith:2018yfb,Chiang:2019ajm,Deppisch:2019ldi,Hayreter:2019dzc,Das:2021esm,Boos:2006xe,Boos:2007eg,Boos:2013cxa}. 

The gauge-Higgs unification (GHU) is one of the approaches to the gauge hierarchy problem~\cite{Hosotani:1983xw, Hosotani:1988bm, Davies:1987ei, Davies:1988wt, Hatanaka:1998yp, Hatanaka:1999sx, Kubo:2001zc, Csaki:2002ur, Burdman:2002se, Scrucca:2003ra, Medina:2007hz}. 
GHU models are constructed in higher dimensional spacetime and the Higgs boson is identified as a part of 
an extra-dimensional component of gauge bosons. 
Hence physics of the Higgs boson is governed by the gauge principle and the Higgs boson mass is generated by quantum corrections in GHU models. 
Many GHU models are proposed for the electroweak unification~\cite{
    Funatsu:2013ni,Funatsu:2014fda,Funatsu:2016uvi,Funatsu:2017nfm,Funatsu:2019ujy,
    Funatsu:2019xwr,Funatsu:2019fry,Funatsu:2020znj,Funatsu:2020haj,Funatsu:2021gnh,
    Matsumoto:2016okl,Hasegawa:2018jze,Yoon:2017cty,Yoon:2018vsc,Kakizaki:2021kof} 
and GHU models for the grand unification are also presented~\cite{
    Hosotani:2015hoa,Hosotani:2017edv,Englert:2019xhz,Englert:2020eep,Lim:2007jv,Kakizaki:2013eba,Kojima:2017qbt,Maru:2019lit,Maru:2019bjr,Angelescu:2021nbp}. 
Among them two types of $SU(3)_C\times SO(5)\times U(1)_X$ GHU models in the warped spacetime have been studied. 
One of them is called the ``A-model", where quark-lepton multiplets are introduced in the vector representation of $SO(5)$~\cite{Funatsu:2013ni,Funatsu:2014fda,Funatsu:2016uvi,Funatsu:2017nfm,Funatsu:2019ujy}. 
The other is called the ``B-model", where quark-lepton multiplets are introduced in the spinor representation of $SO(5)$~\cite{Funatsu:2019xwr,Funatsu:2019fry,Funatsu:2020znj,Funatsu:2020haj,Funatsu:2021gnh}
which has been motivated from the study of the $SO(11)$ gauge-Higgs GUT~\cite{Hosotani:2015hoa,Hosotani:2017edv,Englert:2019xhz,Englert:2020eep},
and is called the GUT inspired GHU model. 

In $SU(3)_C\times SO(5)\times U(1)_X$ GHU models Kaluza-Klein (KK) exited states of the electroweak gauge bosons appear as $W'$ or $Z'$ bosons. 
In our previous work, the constraint $\theta_H\lesssim0.11$ and $m_\text{KK}\gtrsim 8$~TeV has been derived in the A-model from the LHC data at $\sqrt{s} = 8$~TeV, 
where $\theta_H$ is the Aharonov-Bohm (AB) phase in the fifth dimension and $m_\text{KK}$ is the KK mass scale~\cite{Funatsu:2016uvi}. 
The $W$ boson mass ($m_W$) is related to the KK mass scale through $\theta_H$ as $m_W \simeq O(0.1)\times m_\text{KK} \sin\theta_H$~\cite{Funatsu:2019xwr}.
Thus the KK scale can be two orders of magnitude larger than the electroweak scale for $\theta_H\simeq0.1$. 
The effect of $Z'$ bosons is also significant for future linear colliders with polarized electron and positron beams. 
Because of the large asymmetries in the $Z'$ couplings to left- and right-handed fermions, 
some observables in the GHU A-model deviate from those in the SM even at $\sqrt{s}=O(100)$~GeV with the use of polarized beams~\cite{Funatsu:2017nfm,Funatsu:2019ujy}. 
In the $e^+ e^-\to\mu^+\mu^-$ process at the International Linear Collider (ILC) with 
$\sqrt{s}=250$~GeV~\cite{Asner:2013psa,Fujii:2017vwa,Bambade:2019fyw}, 
the deviation of the forward-backward asymmetry from the SM prediction becomes $-2\%$ with the right-handed electron beam for $\theta_H\simeq 0.09$. 
The deviation can be seen with 250 fb$^{-1}$ data. 
The Higgs decay branching ratios are found to be nearly the same as those in the SM~\cite{Funatsu:2013ni}, 
whereas the Higgs triple and quartic couplings deviate $9\%$ and $37\%$ from those in the SM, respectively~\cite{Funatsu:2014fda}. 
It is not easy to distinguish GHU models from the SM from the Higgs data at the ILC. 

The effects of $Z'$ bosons in the GHU B-model at the ILC have been studied in Ref.~\cite{Funatsu:2020haj}. 
The deviation of the forward-backward asymmetry from the SM prediction is about $-1\%$ in the $e^+ e^-\to\mu^+\mu^-$ process at the $250$~GeV ILC with polarized left-handed electron beams for $\theta_H\simeq 0.10$, where the KK mass scale is 13~TeV. 
The deviation of the differential left-right asymmetry reaches to about $-20\%$ in the forward region with the same parameters. 
The A-model and B-model can be distinguished by the dependence in the forward-backward asymmetry and left-right asymmetry 
on the polarization of electron and positron beams, as the two models exhibit opposite dependence on the polarization. 

Another specific feature of the GHU B-model is the appearance of the two step phase transitions at $T\sim 2.6$~TeV and $T=163$~GeV~\cite{Funatsu:2021gnh}. 
At sufficiently high temperature, the effective potential has a minimum at $\theta_H=\pi$. 
The two phases $\theta_H=0$ and $\theta_H=\pi$ become degenerate at $T\sim m_\text{KK}$, where the two phases have $SU(2)_L\times U(1)_{Y}$ and $SU(2)_R\times U(1)_{Y'}$ symmetry, respectively. 
As the temperature becomes lower, the $\theta_H=0$ state becomes the true vacuum and a first-order phase transition from $\theta_H=\pi$ to $\theta_H=0$ takes place at $T \sim 2.6$~TeV. 
This transition is called the left-right phase transition. 
At $T=163$~GeV, the electroweak symmetry breaking (EWSB) occurs and the Higgs boson acquires a vacuum expectation value (VEV). 
This electroweak phase transition is found to be weakly first order. 

In this paper, the $pp \to l\nu$ and $pp \to l^+l^-$ ($l=e,\mu$) processes in the GHU A-model and B-model at the LHC are studied. 
Because significant differences between observables in the SM and those in the GHU models at the ILC are predicted~\cite{Funatsu:2017nfm,Funatsu:2019ujy,Funatsu:2020haj}, 
effects of $W'$ and $Z'$ bosons are expected to be significant at the LHC as well. 
The decay widths of these KK gauge bosons are large because of the large couplings to fermions. 
As a consequence, $W'$ and $Z'$ bosons appear not as narrow peaks but as broad resonances in cross sections. 
Collider signals for these $W'$ and $Z'$ bosons with large decay widths are not studied well, 
and will be studied in this paper. 
Experimental results for the $pp \to l\nu$ and $pp \to l^+l^-$ processes at the LHC at $\sqrt{s}=13$ TeV with up to 140 fb$^{-1}$ data have been published~\cite{ATLAS:2019erb,ATLAS:2020yat,ATLAS:2019lsy,CMS:2021ctt}. 
The constraint on the A-model is updated and is given by $\theta_H\lesssim0.08$ and $m_\text{KK}\gtrsim9.5$~TeV. 
The constraint on the B-model is given by $\theta_H<0.10$ and $m_\text{KK}>13$~TeV. 
At $\sqrt{s}=14$~TeV with the luminosity 300~fb$^{-1}$, 
the discovery significance of the $pp \to e^+e^-$ process in the A-model is 6.49 for $m_\text{KK}=9.5$~TeV, and 
the discovery significance of the $pp \to e\nu$ process in the B-model can be maximally 5.08 for $m_\text{KK}=15$~TeV and can be 1.64 for $m_\text{KK}\simeq18$~TeV. 
At the future High Luminosity LHC (HL-LHC)~\cite{Azzi:2019yne,Cepeda:2019klc,CidVidal:2018eel}, 
the discovery significance of the $pp \to e\nu$ process in the B-model can be maximally 1.61 for $m_\text{KK}\simeq22$~TeV. 

The paper is organized as follows. 
In Sec.~2, the outline of the GHU A-model and B-model is introduced. 
In Sec.~3, definitions of differential cross sections are given. 
In Sec.~4, we evaluate differential cross sections for $pp \to l\nu$ and $pp \to l^+l^-$ processes and constrain the parameters from the experimental results at the LHC Run 2.
In Sec.~5, the predictions for future LHC experiments are shown. 
Section~6 is devoted to a summary and discussions.

\section{Model}
The $SU(3)_C \times SO(5)\times U(1)_X$ GHU A-model and B-model have been given in Ref.~\cite{Funatsu:2014fda} and Ref.~\cite{Funatsu:2019xwr}, in which 
the action, orbifold boundary conditions (BCs), wave functions and formulae to determine the mass spectrum of each field are explained. 
The details of the models are not repeated here. 
In this section, we briefly introduce the models and explain definitions of relevant parameters. 

The $SU(3)_C \times SO(5)\times U(1)_X$ GHU models are defined on the Randall-Sundrum warped spacetime with the metric given by
\begin{align}
 ds^2= g_{MN} dx^M dx^N =e^{-2\sigma(y)} \eta_{\mu\nu}dx^\mu dx^\nu+dy^2,
\end{align} 
where $M,N=0,1,2,3,5$, $\mu,\nu=0,1,2,3$, $y=x^5$,
$\eta_{\mu\nu}=\mbox{diag}(-1,+1,+1,+1)$,
$\sigma(y)=\sigma(y+ 2L)=\sigma(-y)$,
and $\sigma(y)=ky$ for $0 \le y \le L$. 
In terms of the coordinate $z=e^{ky}$ ($1\leq z\leq z_L=e^{kL}$) in the region $0 \leq y \leq L$, 
the metric is written by 
\begin{align}
ds^2= \frac{1}{z^2}
\bigg(\eta_{\mu\nu}dx^{\mu} dx^{\nu} + \frac{dz^2}{k^2}\bigg).
\end{align} 
The bulk region $0<y<L$ ($1<z<z_L$) is an anti-de Sitter (AdS) spacetime. 
The UV and IR branes are located at $y=0$ ($z=1$) and $y=L$ ($z=z_L$), respectively. 
The parameter $k$ is AdS curvature. 
The KK mass scale is given by $m_{\rm KK}\equiv\pi k/(z_L-1) \simeq \pi kz_L^{-1}$ for $z_L\gg 1$. 

The gauge bosons of the $SU(3)_C$, $SO(5)$, and $U(1)_X$ gauge groups are expressed by 
$A_M^{SU(3)_C}$, $A_M^{SO(5)}$, and $A_M^{U(1)_X}$, respectively. 
The BCs for each gauge boson are given by 
\begin{align}
&\begin{pmatrix} A_\mu \cr A_{y} \end{pmatrix} (x,y_j-y) = 
P_{j} \begin{pmatrix} A_\mu \cr - A_{y} \end{pmatrix} (x,y_j+y)P_{j}^{-1}, 
\label{Eq:BC-gauge}
\end{align}
where $(y_0, y_1) = (0, L)$. 
Concretely, 
$P_0=P_1=I_3$ for $A_M^{SU(3)_C}$, $P_0=P_1= 1$ for $A_M^{U(1)_X}$, 
$P_0=P_1=P_{\bf 5}^{SO(5)}=\mbox{diag}\left(I_{4},-I_{1}\right)$ for $A_M^{SO(5)}$ in the vector representation and $P_{\bf 4}^{SO(5)}=\mbox{diag}\left(I_{2},-I_{2}\right)$ in the spinor representation, respectively. 
By the orbifold BCs, $SU(3)_C \times SO(5)\times U(1)_X$ is broken to $SU(3)_C \times SO(4) \times U(1)_X \simeq SU(3)_C \times SU(2)_L \times SU(2)_R \times U(1)_X$. 
$A_\mu^{SU(3)_C}$, $A_\mu^{U(1)_X}$ and $SO(5)/SO(4)$ part of $A_y^{SO(5)}$ have zero modes. 
The zero modes of $A_y^{SO(5)}$ correspond to the Higgs doublet in the SM. 
At this stage the $SO(4) \simeq SU(2)_L \times SU(2)_R$ part of $A_\mu^{SO(5)}$ also have zero modes. 
A brane scalar field $\Phi_{({\bf 1}, {\bf 4})}(x)$ is introduced on the UV brane, which spontaneously breaks $SO(4) \times U(1)_X$ symmetry to $SU(2)_L \times U(1)_Y$ symmetry. 
Finally, $SU(2)_L \times U(1)_Y$ symmetry is dynamically broken to $U(1)_\text{EM}$ symmetry by an AB phase in the fifth dimension. 
Only $\gamma$, $W^\pm$ and $Z$ appear as gauge bosons at low energies. 
At higher energies, in addition to their KK modes $\gamma^{(n)}$, $W^{\pm(n)}$ and $Z^{(n)}$, the KK modes of the broken $SU(2)_R$ gauge bosons, $W_R^{\pm(n)}$ and $Z_R^{(n)}$ can be excited. 
$\gamma^{(n)}$, $Z^{(n)}$ and $Z_R^{(n)}$ appear as  $Z'$ bosons, and $W^{\pm(n)}$ and $W_R^{\pm(n)}$ appear as $W'$ bosons in  $SU(3)_C \times SO(5)\times U(1)_X$ GHU models. 

The 4D Higgs boson doublet $\phi_H(x)$ is the zero mode of $A_z^{SO(5)} = (kz)^{-1} A_y^{SO(5)}$: 
\begin{align}
A_z^{(j5)} (x, z) &= \frac{1}{\sqrt{k}} \, \phi_j (x) u_H (z) + \cdots,\
u_H (z) = \sqrt{ \frac{2}{z_L^2 -1} } \, z ~, \cr
\noalign{\kern 5pt}
\phi_H(x) &= \frac{1}{\sqrt{2}} \begin{pmatrix} \phi_2 + i \phi_1 \cr \phi_4 - i\phi_3 \end{pmatrix} .
\end{align}
Without loss of generality, we set 
$\langle \phi_1 \rangle , \langle \phi_2 \rangle , \langle \phi_3 \rangle =0$ and 
$\langle \phi_4 \rangle \not= 0$, 
which is related to the AB phase $\theta_H$ in the fifth dimension by 
$\langle \phi_4 \rangle = \theta_H f_H$, where 
\begin{align}
&f_H = \frac{2}{g_w} \sqrt{ \frac{k}{L(z_L^2 -1)}} ~.
\label{fH1}
\end{align}

The representations and BCs of the matter fields are different between the A-model and B-model, 
where the matter fields are introduced both in the bulk and on the UV brane. 
The matter fields of these two models are listed in Table~\ref{Tab:matter}. 
In the A-model, the SM quarks and leptons are identified with the zero modes of the $SO(5)$-vector fermions $\Psi^\alpha_a$ ($a=1,..., 4$ and $\alpha=1,2,3$). 
The BCs for those fields are given by 
\begin{align}
    \Psi^\alpha_a (x, y_j-y)= &~ P_{\bf 5}^{SO(5)} \Gamma^5 \Psi^\alpha_a (x, y_j+y)~.\label{quarkBC1}
\end{align}
Meanwhile, 
in the B-model, the SM quarks and leptons are identified with the zero modes of the $SO(5)$-spinor and singlet fermions 
$\Psi_{({\bf 3,4})}^{\alpha}$,
$\Psi_{({\bf 3,1})}^{\pm \alpha}$, and 
$\Psi_{({\bf 1,4})}^{\alpha}$ $(\alpha=1,2,3)$.
These fields obey the following BCs: 
\begin{align}
&\Psi_{({\bf 3,4})}^{\alpha} (x, y_j - y) = 
 - P_{\bf 4}^{SO(5)} \gamma^5 \Psi_{({\bf 3,4})}^{\alpha} (x, y_j + y),
 \nonumber\\
&\Psi_{({\bf 3,1})}^{\pm \alpha} (x, y_j - y) =
\mp \gamma^5 \Psi_{({\bf 3,1})}^{\pm \alpha} (x, y_j + y),\nonumber\\
&\Psi_{({\bf 1,4})}^{\alpha} (x, y_j - y) = 
 - P_{\bf 4}^{SO(5)} \gamma^5 \Psi_{({\bf 1,4})}^{\alpha} (x, y_j + y).
\label{quarkBC2}
\end{align}
With BCs~(\ref{quarkBC2}), the $SU(2)_L$ ($SU(2)_R$) components of $\Psi_{({\bf 3,4})}^{\alpha}$ and $\Psi_{({\bf 1,4})}^{\alpha}$ have the left-(right-)handed zero modes 
and $\Psi_{({\bf 3,1})}^{+ \alpha}$ ($\Psi_{({\bf 3,1})}^{- \alpha}$) has the left-(right-)handed zero modes, respectively. 
These zero modes mix with each other through the AB phase $\theta_H$ and brane interactions. 

There are matter fields having no zero modes, which is referred as the dark fermions. 
Those dark fermions are necessary to have dynamical EWSB with suitable value of $\theta_H$. 
Dark fermions do not couple directly with the quarks and leptons and the lightest modes of dark fermions have masses about $m_\text{KK}/2$. 
Decay  of the first KK gauge bosons to dark fermions are either forbidden or negligible. 
The details of the dark fermions are given in Ref.~\cite{Funatsu:2019xwr}. 

\begin{table}[tbh]
    \renewcommand{\arraystretch}{1.2}
    \centering
    \caption{Matter fields. Brane fields and the symmetry at the UV brane are also shown. }
    \vskip 10pt
    \begin{tabular}{|c|c|c|}
    \hline
    & A-model & B-model \\
    \hline \hline
    quark
    &$\Psi^{\alpha}_1:(\bm{3}, \bm{5})_{+\frac{2}{3}}, ~ \Psi^{\alpha}_2:(\bm{3}, \bm{5})_{-\frac{1}{3}}$ 
    &$\Psi^{\alpha}_{({\bf 3,4})}:(\bm{3}, \bm{4})_{+\frac{1}{6}}, ~ \Psi^{\pm \alpha}_{({\bf 3,1})}:(\bm{3}, \bm{1})_{-\frac{1}{3}}^\pm$\\
    lepton
    &$\Psi^{\alpha}_3:(\bm{1}, \bm{5})_{0}, ~ \Psi^{\alpha}_4:(\bm{1}, \bm{5})_{-1}$
    &$\Psi^{\alpha}_{({\bf 1,4})}(\bm{1}, \bm{4})_{-\frac{1}{2}}$ \rule[-3mm]{0mm}{8mm}\\
    \hline
    dark fermion 
    &$\Psi_F^\delta:(\bm{1}, \bm{4})_{+\frac{1}{2}}$ 
    &$\Psi_{F_q}^\beta:(\bm{3}, \bm{4})_{+\frac{1}{6}}, ~ \Psi_{F_l}^\beta:(\bm{1}, \bm{4})_{-\frac{1}{2}}$ \rule[-3mm]{0mm}{8mm}\\
    &
    &$\Psi_V^{\pm\gamma}: (\bm{1}, \bm{5})_{0}^\pm$\\
    \hline \hline
    brane fermion  
    &$\begin{matrix} 
    \hat{\chi}^q_{1,2,3 R}:(\bm{3}, [\bm{2,1}])_{+\frac{7}{6}, +\frac{1}{6}, -\frac{5}{6}} \cr
    \hat{\chi}^q_{1,2,3 R}:(\bm{1}, [\bm{2,1}])_{+\frac{1}{2}, -\frac{1}{2}, -\frac{3}{2}} \end{matrix}$ \rule[-7.5mm]{0mm}{16mm}
    &$\chi:(\bm{1}, \bm{1})_{0}$\\
    \hline
    brane scalar 
    &$\hat{\Phi}:(\bm{1}, [\bm{1,2}])_{+\frac{1}{2}}$ 
    &$\Phi_{(\bm{1}, \bm{4})}:(\bm{1}, \bm{4})_{\frac{1}{2}}$ \rule[-3mm]{0mm}{8mm} \\
    \hline
    $\begin{matrix} \text{symmetry ~of} \cr \text{brane ~interactions} \end{matrix}$
    &$SU(3)_C \times SO(4) \times U(1)_X$ &$SU(3)_C \times SO(5) \times U(1)_X$ \\
    \hline
    \end{tabular}
    \label{Tab:matter}
\end{table}

The procedure to determine the model parameters is explained in Refs~\cite{Funatsu:2014fda, Funatsu:2019xwr, Funatsu:2019fry, Funatsu:2020znj}. 
With specified parameters, masses and couplings are all determined. 
It has been shown that one can introduce the Cabibbo-Kobayashi-Maskawa (CKM) matrix in the GHU B-model while flavor changing neutral currents (FCNCs) are naturally suppressed~\cite{Funatsu:2019fry}. 
The tree-level FCNCs exist only in the down-type quark sector and the magnitude of their couplings is of the order of $O(10^{-6})$. 
In this paper, the CKM matrix is not introduced for simplicity. 

As benchmark points, eight parameter sets are taken as shown in Table~\ref{Table:Mass-Width-Vector-Bosons}. 
These benchmark points are chosen for the following reason. 
In the A-model, there are two free parameters $z_L$ (or $m_\text{KK}$) and $n_F$ (the number of dark fermions). 
The effects of $n_F$ on physics of the gauge bosons, Higgs bosons, quarks and leptons turn out very small. 
Once a set of parameters ($n_F$, $z_L$) is set, $\theta_H$ is determined and couplings among these fields are determined by $\theta_H$. 
Hence a relevant free parameter is effectively $\theta_H$ only. 
The value of $n_F$ affects a lower limit of $\theta_H$ and a dark fermion mass. 
For a larger $n_F$, the lower limit of $\theta_H$ is smaller and the lowest mode of the dark fermion has a lower mass. 
We take $n_F=4$, where the lower limit of $\theta_H$ is $\theta_H\simeq0.08$. 
The top quark and Higgs boson mass cannot be reproduced for $n_F=4$ and $\theta_H<0.08$ in the A-model. 
In the B-model, there are four free parameters in the dark fermion sector, and $z_L$ and $\theta_H$ are not uniquely determined. 
These parameters are constrained to reproduce the top quark mass and EWSB. 
The top quark mass is realized only for $z_L\gtrsim 10^{8.1}$, and the EWSB is triggered only for $z_L\lesssim 10^{15.5}$ for $\theta_H= 0.10$. 
Consistent parameter sets are obtained only for $10^{8.1}\lesssim z_L\lesssim 10^{15.5}$~\cite{Funatsu:2020znj}. 
We set $\theta_H=0.10$ and choose, as typical values, integral values of $m_\text{KK}$/TeV in the allowed region. 
To check the $\theta_H$-dependence, $\theta_H=0.11,\ 0.10,\ 0.09$ are chosen for $m_\text{KK} = 13$~TeV. 

\begin{table}[thb]{\small
    \centering
    \begin{tabular}{l|cc|cccccc|c}
    \hline
     Name&$\theta_H$&$m_\text{KK}$&$z_L$&$k$
     &$m_{W^{(1)}}$&$\Gamma_{W^{(1)}}$
     &$m_{W_R^{(1)}}$&$\Gamma_{W_R^{(1)}}$
     &Table\\
     &\mbox{[rad]}&[TeV]&&[GeV]&[TeV]&[TeV]&[TeV]&[TeV]
     &\\ 
    \hline 
     \hspace{0.5em}A1 & 0.10 & 8.063 & 2.900$\times10^{4}$ & 7.443$\times10^{7}$ & 6.585 & 0.236 & 6.172 & 0.098 &\\
     \hspace{0.5em}A2 & 0.09 & 8.721 & 1.700$\times10^{4}$ & 4.719$\times10^{7}$ & 7.149 & 0.271 & 6.676 & 0.101 &\ref{Table:Couplings-Wprime_A}\\
     \hspace{0.5em}A3 & 0.08 & 9.544 & 1.010$\times10^{4}$ & 3.068$\times10^{7}$ & 7.855 & 0.356 & 7.305 & 0.105 &\\
    \hline
    \hline 
     \hspace{0.5em}B$^{\rm L}$&0.10&11&1.980$\times10^{8\ }$&6.933$\times10^{11}$&8.713&5.925&8.420&0.218&\ref{Table:Couplings-Wprime_thetaH=010-mKK=11}\\
     \hspace{0.5em}B          &0.10&13&3.865$\times10^{11}$ &1.599$\times10^{15}$&10.20&9.812&9.951&0.368&\ref{Table:Couplings-Wprime_thetaH=010-mKK=13}\\
     \hspace{0.5em}B$^{\rm H}$&0.10&15&2.667$\times10^{15}$ &1.273$\times10^{19}$&11.69&14.85&11.48&0.565&\ref{Table:Couplings-Wprime_thetaH=010-mKK=15}\\
    \hline
    \hline 
     \hspace{0.5em}B$^+$&0.11&13&1.021$\times10^{14}$ &4.223$\times10^{17}$&10.15&11.75&9.951&0.445&\ref{Table:Couplings-Wprime_thetaH=011-mKK=13}\\
     \hspace{0.5em}B    &0.10&13&3.865$\times10^{11}$ &1.599$\times10^{15}$&10.20&9.812&9.951&0.368&\ref{Table:Couplings-Wprime_thetaH=010-mKK=13}\\
     \hspace{0.5em}B$^-$&0.09&13&2.470$\times10^{9\ }$&1.022$\times10^{13}$&10.26&7.993&9.951&0.291&\ref{Table:Couplings-Wprime_thetaH=009-mKK=13}\\
    \hline
    \end{tabular}
     \caption{
     Masses and widths of $W^{(1)}$ and $W_R^{(1)}$ are listed. 
     Those in the A-model are shown for $\theta_H=0.10$, $0.09$, $0.08$ in the upper table. 
     Those in the B-model are shown for $\theta_H=0.10$ and three $m_{\rm KK}=11,\ 13,\ 15$~TeV in the middle table and for $m_{\rm KK}=13$~TeV and three $\theta_H=0.11,\ 0.10,\ 0.09$ in the lower table.
     }
    \label{Table:Mass-Width-Vector-Bosons}}
\end{table}

\begin{table}[htb]
    \centering
        \begin{tabular}{c|cc|cc|cc}
        \hline
         & \multicolumn{2}{|c|}{A1} & \multicolumn{2}{|c|}{A2} & \multicolumn{2}{|c}{A3}\\
         $ff'$ &$g_{W^{(1)}ff'}^L$ & $g_{W^{(1)}ff'}^R$ & $g_{W^{(1)}ff'}^L$ & $g_{W^{(1)}ff'}^R$ & $g_{W^{(1)}ff'}^L$ & $g_{W^{(1)}ff'}^R$ \\
         \hline
         $e \nu_e$     & $-0.3675$ & 0 & $-0.3785$ & 0 & $-0.3901$ & 0 \\
         $\mu \nu_\mu$ & $-0.3675$ & 0 & $-0.3785$ & 0 & $-0.3901$ & 0 \\
         $\tau \nu_\tau$&$-0.3670$ & 0 & $-0.3779$ & 0 & $-0.3898$ & 0 \\
         $ud$ & $-0.3675$ & 0 & $-0.3785$ & 0 & $-0.3901$ & 0 \\
         $cs$ & $-0.3675$ & 0 & $-0.3785$ & 0 & $-0.3901$ & 0 \\
         $tb$ & $+1.4588$ & 0 & $+1.5635$ & 0 & $+1.8313$ & 0 \\
        \hline
        \end{tabular}
     \caption{
     Coupling constants of 1st KK $W$ boson to fermions in units of $g_w/\sqrt{2}$ are listed for A-model,
     where $\sin^2\theta_W^0=0.23126$. 
     The value less than $10^{-4}$ is written as $0$.
     The $W$ boson couplings are $g_{Wff'}^L=1.000$ for $ff'\neq tb$ and 
     $g_{Wtb}^L=0.9993$ for A1, $g_{Wtb}^L=0.9994$ for A2 and A3. 
     The couplings of the first KK $W_R$ boson to fermions are exactly zero. 
     }
    \label{Table:Couplings-Wprime_A}
\end{table}

\begin{table}[htb]
    \centering
        \begin{tabular}{ccccccc}
        \hline
         $ff'$ &$g_{Wff'}^L$ & $g_{Wff'}^R$ & $g_{W^{(1)}ff'}^L$ & $g_{W^{(1)}ff'}^R$ & $g_{W_R^{(1)}ff'}^L$ & $g_{W_R^{(1)}ff'}^R$\\
         \hline
         $e \nu_e$     & 0.9976 & 0 & 5.7451 & 0 & 0.0146 & 0 \\
         $\mu \nu_\mu$ & 0.9976 & 0 & 5.4705 & 0 & 0.0139 & 0 \\
         $\tau \nu_\tau$&0.9976 & 0 & 5.2877 & 0 & 0.0134 & 0 \\
         $ud$ & 0.9976 & 0 & 5.5626 & 0 & 0.0141 & 0 \\
         $cs$ & 0.9976 & 0 & 5.3588 & 0 & 0.0136 & 0 \\
         $tb$ & 0.9980 & 0 & 4.4108 & 0 & 0.0113 & $-$0.0344 \\
        \hline
        \end{tabular}
     \caption{
     Coupling constants of charged vector bosons, $W'$ bosons, to fermions in units of $g_w/\sqrt{2}$
     are listed for $\theta_H=0.10$ and $m_\text{KK}=13$~TeV (B) in Table~\ref{Table:Mass-Width-Vector-Bosons},
     where $\sin^2\theta_W^0=0.2306$. 
     The value less than $10^{-4}$ is written as $0$.
     }
    \label{Table:Couplings-Wprime_thetaH=010-mKK=13}
\end{table}

\begin{table}[htb]
    \centering
        \begin{tabular}{ccccccc}
        \hline
         $ff'$ &$g_{Wff'}^L$ & $g_{Wff'}^R$ & $g_{W^{(1)}ff'}^L$ & $g_{W^{(1)}ff'}^R$ & $g_{W_R^{(1)}ff'}^L$ & $g_{W_R^{(1)}ff'}^R$\\
         \hline
         $e \nu_e$     & 0.9977 & 0 & 5.0203 & 0 & 0.0127 & 0 \\
         $\mu \nu_\mu$ & 0.9977 & 0 & 4.7423 & 0 & 0.0121 & 0 \\
         $\tau \nu_\tau$&0.9977 & 0 & 4.5451 & 0 & 0.0116 & 0 \\
         $ud$ & 0.9977 & 0 & 4.8380 & 0 & 0.0123 & 0 \\
         $cs$ & 0.9977 & 0 & 4.6229 & 0 & 0.0118 & 0 \\
         $tb$ & 0.9982 & 0 & 3.1365 & 0 & 0.0082 & $-$0.0455 \\
        \hline
        \end{tabular}
     \caption{
     Coupling constants of charged vector bosons, $W'$ bosons, to fermions in units of $g_w/\sqrt{2}$
     are listed for $\theta_H=0.10$ and $m_\text{KK}=11$~TeV (B$^{\rm L}$) in Table~\ref{Table:Mass-Width-Vector-Bosons},
     where $\sin^2\theta_W^0=0.2306$.
     Other information is the same as in Table~\ref{Table:Couplings-Wprime_thetaH=010-mKK=13}.
     }
    \label{Table:Couplings-Wprime_thetaH=010-mKK=11}
\vspace{1em}
    \centering
        \begin{tabular}{ccccccc}
        \hline
         $ff'$ &$g_{Wff'}^L$ & $g_{Wff'}^R$ & $g_{W^{(1)}ff'}^L$ & $g_{W^{(1)}ff'}^R$ & $g_{W_R^{(1)}ff'}^L$ & $g_{W_R^{(1)}ff'}^R$\\
         \hline
         $e \nu_e$     & 0.9976 & 0 & 6.4691 & 0 & 0.0164 & 0 \\
         $\mu \nu_\mu$ & 0.9976 & 0 & 6.2055 & 0 & 0.0157 & 0 \\
         $\tau \nu_\tau$&0.9976 & 0 & 6.0376 & 0 & 0.0153 & 0 \\
         $ud$ & 0.9976 & 0 & 6.2923 & 0 & 0.0159 & 0 \\
         $cs$ & 0.9976 & 0 & 6.1021 & 0 & 0.0155 & 0 \\
         $tb$ & 0.9978 & 0 & 5.3389 & 0 & 0.0136 & $-$0.0294 \\
        \hline
        \end{tabular}
     \caption{
     Coupling constants of charged vector bosons, $W'$ bosons, to fermions in units of $g_w/\sqrt{2}$
     are listed for $\theta_H=0.10$ and $m_\text{KK}=15$~TeV (B$^{\rm H}$) in Table~\ref{Table:Mass-Width-Vector-Bosons},
     where $\sin^2\theta_W^0=0.2306$. 
     Other information is the same as in Table~\ref{Table:Couplings-Wprime_thetaH=010-mKK=13}.
     }
    \label{Table:Couplings-Wprime_thetaH=010-mKK=15}
\end{table}

\begin{table}[htb]
    \centering
        \begin{tabular}{ccccccc}
        \hline
         $ff'$ &$g_{Wff'}^L$ & $g_{Wff'}^R$ & $g_{W^{(1)}ff'}^L$ & $g_{W^{(1)}ff'}^R$ & $g_{W_R^{(1)}ff'}^L$ & $g_{W_R^{(1)}ff'}^R$\\
         \hline
         $e \nu_e$     & 0.9971 & 0 & 6.2134 & 0 & 0.0190 & 0 \\
         $\mu \nu_\mu$ & 0.9971 & 0 & 5.9455 & 0 & 0.0182 & 0 \\
         $\tau \nu_\tau$&0.9971 & 0 & 5.7724 & 0 & 0.0177 & 0 \\
         $ud$ & 0.9971 & 0 & 6.0342 & 0 & 0.0185 & 0 \\
         $cs$ & 0.9971 & 0 & 5.8391 & 0 & 0.0179 & 0 \\
         $tb$ & 0.9974 & 0 & 5.0226 & 0 & 0.0155 & $-$0.0309 \\
        \hline
        \end{tabular}
     \caption{
     Coupling constants of charged vector bosons, $W'$ bosons, to fermions in units of $g_w/\sqrt{2}$
     are listed for $\theta_H=0.11$ and $m_\text{KK}=13$~TeV (B$^+$) in Table~\ref{Table:Mass-Width-Vector-Bosons},
     where $\sin^2\theta_W^0=0.2305$. 
     Other information is the same as in
     Table~\ref{Table:Couplings-Wprime_thetaH=010-mKK=13}.
     }
    \label{Table:Couplings-Wprime_thetaH=011-mKK=13}
\vspace{1em}
    \centering
        \begin{tabular}{ccccccc}
        \hline
         $ff'$ &$g_{Wff'}^L$ & $g_{Wff'}^R$ & $g_{W^{(1)}ff'}^L$ & $g_{W^{(1)}ff'}^R$ & $g_{W_R^{(1)}ff'}^L$ & $g_{W_R^{(1)}ff'}^R$\\
         \hline
         $e \nu_e$     & 0.9981 & 0 & 5.2761 & 0 & 0.0108 & 0 \\
         $\mu \nu_\mu$ & 0.9981 & 0 & 4.9979 & 0 & 0.0103 & 0 \\
         $\tau \nu_\tau$&0.9981 & 0 & 4.8056 & 0 & 0.0099 & 0 \\
         $ud$ & 0.9981 & 0 & 5.0926 & 0 & 0.0105 & 0 \\
         $cs$ & 0.9981 & 0 & 4.8810 & 0 & 0.0101 & 0 \\
         $tb$ & 0.9985 & 0 & 3.7015 & 0 & 0.0078 & $-$0.0397 \\
        \hline
        \end{tabular}
     \caption{
     Coupling constants of charged vector bosons, $W'$ bosons, to fermions in units of $g_w/\sqrt{2}$
     are listed for $\theta_H=0.09$ and $m_\text{KK}=13$~TeV (B$^-$) in Table~\ref{Table:Mass-Width-Vector-Bosons},
     where $\sin^2\theta_W^0=0.2307$. 
     Other information is the same as in
     Table~\ref{Table:Couplings-Wprime_thetaH=010-mKK=13}.
     }
    \label{Table:Couplings-Wprime_thetaH=009-mKK=13}
\end{table}

The masses and decay widths of the $W^{(1)}$ and $W_R^{(1)}$ bosons are shown in Table~\ref{Table:Mass-Width-Vector-Bosons}, where the decay widths are calculated by the couplings shown in 
Tables \ref{Table:Couplings-Wprime_A}, \ref{Table:Couplings-Wprime_thetaH=010-mKK=13}, \ref{Table:Couplings-Wprime_thetaH=010-mKK=11}, \ref{Table:Couplings-Wprime_thetaH=010-mKK=15}, 
\ref{Table:Couplings-Wprime_thetaH=011-mKK=13} and \ref{Table:Couplings-Wprime_thetaH=009-mKK=13}. 
Those of the $\gamma^{(1)}$, $Z^{(1)}$ and $Z_R^{(1)}$ bosons are shown in Refs.~\cite{Funatsu:2019ujy,Funatsu:2020haj}. 
The masses of $W^{(1)}$, $Z^{(1)}$ and $\gamma^{(1)}$ bosons are almost degenerate and about 0.8 times the KK mass scale. 
The masses of $W_R^{(1)}$ and $Z_R^{(1)}$ bosons are slightly lighter than those of $W^{(1)}$ and $Z^{(1)}$ bosons. 

The couplings among particles depend on the behavior of their wave-functions. 
In the $\theta_H\to 0$ limit,  $W$ and $W_R$ bosons are purely  $SU(2)_L$ and $SU(2)_R$ gauge bosons. 
In the A-model, zero modes of left-handed quarks and leptons are $SU(2)_L \times SU(2)_R$ bidoublets, 
whereas right-handed quarks and leptons are singlets. 
$W$ and $W_R$ bosons do not couple with  right-handed quarks and leptons, 
as those couplings are determined by overlap integrals of their wave-functions. 
$W^{(1)}$ and $W_R^{(1)}$ bosons are localized near the IR brane and 
zero modes of left-handed quarks and leptons are localized near the UV brane except for the top quark. 
Consequently, $W^{(1)}$ couples with left-handed SM fermions very weakly except for the top quark and does not couple with right-handed SM fermions. 
The decay width of $W^{(1)}$ is narrow as shown in Table~\ref{Table:Mass-Width-Vector-Bosons}. 

On the other hand, the $W^{(1)}$ couplings with left-handed fermions are large in the B-model. 
The zero modes of the left- and right-handed quarks and leptons in the B-model are purely $SU(2)_L$ and $SU(2)_R$ components, respectively. 
Therefore $W^{(1)}$ mainly couples with left-handed quarks and leptons and $W_R^{(1)}$ mainly couples with right-handed ones for small $\theta_H$. 
In contrast to the A-model, 
the zero modes of left-handed quarks and leptons are localized near the IR brane and 
the zero modes of right-handed quarks and leptons are localized near the UV brane in the B-model. 
Hence, the $W^{(1)}$ couplings with left-handed fermions are large, numerically being of $O(1)\times g_w/\sqrt{2}$. 
In contrast to it, the $W_R^{(1)}$ couplings with left-handed fermions are small, numerically being of $O(10^{-2})\times g_w/\sqrt{2}$. 
The $W^{(1)}$ and $W_R^{(1)}$ couplings with right-handed fermions are tiny and negligible except for $W_R^{(1)}\bar{t}_R\:b_R$ coupling. 
Because of the large $W^{(1)}$ couplings with left-handed fermions, the decay width of $W^{(1)}$ becomes very wide, 
which is numerically $\Gamma_{W^{(1)}}/m_{W^{(1)}}=0.68$ -- $1.27$ as shown in Table~\ref{Table:Mass-Width-Vector-Bosons}. 

The contribution from the higher KK modes are calculated in Ref.~\cite{Funatsu:2020haj} and found to be small. 
Numerically, the deviation of the $e^+e^-\to \mu^+\mu^-$ cross section with the first and second KK modes 
from the deviation  with only the first KK modes is $O(1)$\% for $\sqrt{s} < 3$ TeV. 
Thus we take only the first KK modes into account in the following.

\section{Differential Cross Sections}
In this section, formulae for the cross sections of the $pp\to l^+l^-$ and $pp\to l\nu$ processes are summarized~\cite{Halzen:1984mc,Peskin:1995ev}. 

The cross section of the $pp\to l^+l^-$ process, $\sigma_{pp\to l^+l^-}$, is written in terms of the parton-level cross section $\sigma_{f\bar{f}\to l^+l^-}(s)$ as 
\begin{align}
    \sigma_{pp\to l^+l^-}
    &=\sum_f\int^1_0 dx_1 \int^1_0 dx_2 \:\sigma_{f\bar{f}\to l^+l^-}(s) \left(F_f(x_1, Q)F_{\bar{f}}(x_2, Q)+F_f(x_2, Q)F_{\bar{f}}(x_1, Q)\right) \nonumber\\
    &=\sum_f\int^{s_{pp}}_0 ds \int^1_{s/s_{pp}} dx_1 \frac{2}{x_1 s_{pp}} \sigma_{f\bar{f}\to l^+l^-}(s) F_f(x_1, \sqrt{s})F_{\bar{f}}\left(\frac{s}{x_1s_{pp}}, \sqrt{s}\right),
\end{align}
where $F_{f}(x_{1,2}, Q)$ are the parton distribution function (PDF) at an energy scale $Q$ and we take $Q=\sqrt{s}$ in this paper. 
By introducing the invariant mass of the lepton pair denoted as $m_{ll}\equiv \sqrt{s}$, 
the differential cross section with respect to the invariant mass is written as 
\begin{align}
    \frac{d\sigma_{pp\to l^+l^-}}{dm_{ll}} &=\sum_f\int^1_{m_{ll}^2/s_{pp}} dx_1 \frac{4m_{ll}}{x_1 s_{pp}} \sigma_{f\bar{f}\to l^+l^-}(s=m_{ll}^2) F_f(x_1, m_{ll})F_{\bar{f}}\left(\frac{m_{ll}^2}{x_1s_{pp}}, m_{ll}\right).
\end{align}

Similarly, the cross section of the $pp\to l\nu$ process, $\sigma_{pp\to l\nu}$, is written as 
\begin{align}
    \sigma_{pp\to l\nu}
    =\sum_{f,\ f'}\int^{s_{pp}}_0 ds \int^1_{-1} d\cos\theta \int^1_{s/s_{pp}} dx_1 
    \frac{2}{x_1 s_{pp}} \frac{d\sigma_{f\bar{f'}\to l\nu}(s)}{d\cos\theta} F_f(x_1, \sqrt{s})F_{\bar{f}'}\left(\frac{s}{x_1s_{pp}}, \sqrt{s}\right),
\end{align}
where $\sigma_{f\bar{f'}\to l\nu}(s)$ is the parton-level cross section. 
The cross section of the $pp\to l\nu$ process is explored by measuring the transverse mass ($m_\text{T}$) or the transverse momentum of the charged-lepton ($p_\text{T}$). 
These two parameters are related as $m_\text{T}=\sqrt{2p_\text{T}E_\text{T}^\text{miss}(1-\cos\phi)}$, 
where $E_\text{T}^\text{miss}$ is the missing energy and $\phi$ is the angle between the charged-lepton and missing transverse momentum in the transverse plane. 
For $pp$ collisions at high energies, the transverse momentum of the parton may be ignored. 
$\phi\simeq\pi$ and the masses of the leptons are ignored, and one finds $m_\text{T}\simeq 2\, p_\text{T}$. 
The transverse momentum is defined as $p_\text{T}\equiv (\sqrt{s}/2)\vert\sin\theta\vert$. 
The total cross section is written by using the transverse mass as 
\begin{align}
    \sigma_{pp\to l\nu}
    &=\sum_{f,\ f'}\int^{\sqrt{s_{pp}}}_0 dm_{ll} \int^{m_{ll}}_0 dm_\text{T} \int^1_{\frac{m_{ll}^2}{s_{pp}}} dx_1 
    \frac{8m_\text{T}}{x_1 s_{pp} m_{ll}\sqrt{1-\frac{m_\text{T}^2}{m_{ll}^2}}} 
    F_f(x_1, m_{ll})F_{\bar{f}'}\left(\frac{m_{ll}^2}{x_1s_{pp}}, m_{ll}\right)\nonumber\\
    &\qquad\qquad\times \left(\left.\frac{d\sigma_{f\bar{f'}\to l\nu}(s)}{d\cos\theta}\right|_{\cos\theta=+\sqrt{1-\frac{m_\text{T}^2}{m_{ll}^2}}}
    +\left.\frac{d\sigma_{f\bar{f'}\to l\nu}(s)}{d\cos\theta}\right|_{\cos\theta=-\sqrt{1-\frac{m_\text{T}^2}{m_{ll}^2}}}\right) \nonumber\\
    &=\sum_{f,\ f'}\int^{\sqrt{s_{pp}}}_0 dm_\text{T} \int^{\sqrt{s_{pp}}}_{m_\text{T}} dm_{ll} \int^1_{\frac{m_{ll}^2}{s_{pp}}} dx_1 
    \frac{8m_\text{T}}{x_1 s_{pp} m_{ll}\sqrt{1-\frac{m_\text{T}^2}{m_{ll}^2}}} 
    F_f(x_1, m_{ll})F_{\bar{f}'}\left(\frac{m_{ll}^2}{x_1s_{pp}}, m_{ll}\right)\nonumber\\
    &\qquad\qquad\times \left(\left.\frac{d\sigma_{f\bar{f'}\to l\nu}(s)}{d\cos\theta}\right|_{\cos\theta=+\sqrt{1-\frac{m_\text{T}^2}{m_{ll}^2}}}
    +\left.\frac{d\sigma_{f\bar{f'}\to l\nu}(s)}{d\cos\theta}\right|_{\cos\theta=-\sqrt{1-\frac{m_\text{T}^2}{m_{ll}^2}}}\right) .
\end{align}
The differential cross section with respect to the transverse mass is 
\begin{align}
    \frac{d\sigma_{pp\to l\nu}}{dm_\text{T}}
    &=\sum_{f,\ f'}\int^{\sqrt{s_{pp}}}_{m_\text{T}} dm_{ll} \int^1_{\frac{m_{ll}^2}{s_{pp}}} dx_1 
    \frac{8m_\text{T}}{x_1 s_{pp} m_{ll}\sqrt{1-\frac{m_\text{T}^2}{m_{ll}^2}}} 
    F_f(x_1, m_{ll})F_{\bar{f}'}\left(\frac{m_{ll}^2}{x_1s_{pp}}, m_{ll}\right)\nonumber\\
    &\qquad\qquad\times \left(\left.\frac{d\sigma_{f\bar{f'}\to l\nu}(s)}{d\cos\theta}\right|_{\cos\theta=+\sqrt{1-\frac{m_\text{T}^2}{m_{ll}^2}}}
    +\left.\frac{d\sigma_{f\bar{f'}\to l\nu}(s)}{d\cos\theta}\right|_{\cos\theta=-\sqrt{1-\frac{m_\text{T}^2}{m_{ll}^2}}}\right) .
\end{align}

\section{Constraints from LHC Experiments}
Constraints on the GHU A-model from the early stage of LHC experiment are estimated in Refs~\cite{Funatsu:2014fda,Funatsu:2016uvi}. 
In this section, we update the constraints by using the LHC run 2 data. 
Constraints on the GHU B-model are obtained as well. 
We use CT10~\cite{Lai:2010vv} for the PDF and ManeParse~\cite{Clark:2016jgm} to numerically evaluate the cross sections. 

\begin{figure}[thb]
    \centering
    \subfigure{\includegraphics[width=0.495\columnwidth]{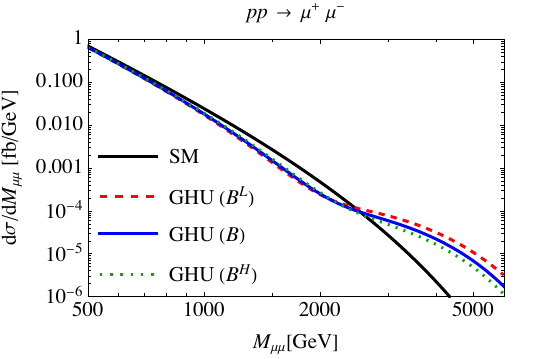}}
    \subfigure{\includegraphics[width=0.495\columnwidth]{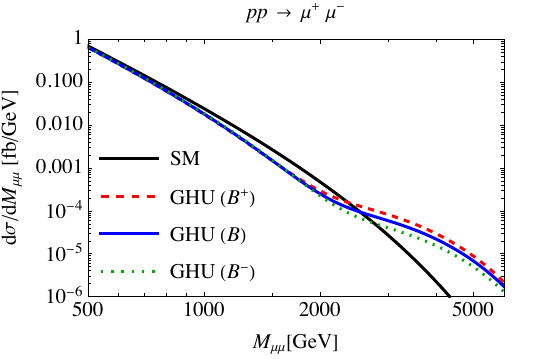}}
     \caption{
     Differential cross sections $d\sigma(pp\to{\mu^+\mu^-})/d M_{\mu\mu}$ for $\sqrt{s}=13$~TeV. 
     The differential cross section in the SM is shown by the black solid line. 
     In the left figure, the differential cross sections with fixed $\theta_H=0.10$ and $m_\text{KK}=11$~TeV (B$^\text{L}$), 13~TeV (B), 15~TeV (B$^\text{H}$) are shown. 
     In the right figure, the differential cross sections with fixed $m_\text{KK}=13$~TeV and $\theta_H=0.11$ (B$^+$), $\theta_H=0.10$ (B), $\theta_H=0.09$ (B$^-$) are shown. 
     }
    \label{Figure:sigma-mumu}
\end{figure}

First, we recall the constraints on the $Z'$ bosons from the $pp\to e^+ e^-$ and $\mu^+\mu^-$ processes. 
To see the $\theta_H$ and $m_\text{KK}$ dependence, the differential cross sections $d\sigma(pp\to{\mu^+\mu^-})/d M_{\mu\mu}$ are shown 
for the parameter sets (B$^\text{L}$, B, B$^\text{H}$) and (B$^+$, B, B$^-$) in Fig.~\ref{Figure:sigma-mumu}. 
The decay widths of the $Z'$ bosons are large, 
and therefore we refer the constraint from the non-resonant searches in the dilepton final states at the ATLAS group~\cite{ATLAS:2020yat}. 
The $d\sigma(pp\to{l^+l^-})/d M_{ll}$ in the GHU model is $O(1)$\% smaller for $M_{ll}\simeq500$~GeV and 20--30\% smaller for $M_{ll}\simeq1000$~GeV, 
and becomes larger for $M_{ll}\gtrsim2500$~GeV than that in the SM as shown in Fig.~\ref{Figure:sigma-mumu}. 
Hence we follow the analysis in the ``destructive interference" case in Ref.~\cite{ATLAS:2020yat}. 
For simplicity, we assume that the acceptance and efficiency are independent of the invariant mass. 
Therefore the acceptance times efficiency are introduced as constant factors of the differential cross sections 
and are determined to minimize the $\chi^2$ values in the control regions (CRs). 
The number of events is obtained by integrating the differential cross sections over the signal regions (SRs). 
The CRs and SRs for each process are 
CR: [310,~1450]~GeV and SR: [2770,~6000]~GeV for $pp\to e^+ e^-$ and
CR: [320,~1250]~GeV and SR: [2570,~6000]~GeV for $pp\to \mu^+ \mu^-$, respectively~\cite{ATLAS:2020yat}. 
$N_\text{GHU}$ and $N_\text{SM}$ denote the number of events in the GHU model and that in the SM, respectively. 
The number of signal events $N_\text{sig}$ is defined as the excess of $N_\text{GHU}$ from $N_\text{SM}$; 
$N_\text{sig}\equiv N_\text{GHU}- N_\text{SM}$. 
$N_\text{SM}=1.2$ for $pp\to e^+ e^-$ and $N_\text{SM}=1.8$ for $pp\to \mu^+ \mu^-$. 
$N_\text{sig}$ for the parameter sets (A1, A2, A3) are (21, 10, 4.9) for $pp\to e^+ e^-$ and (18, 8.6, 4.0) for $pp\to \mu^+ \mu^-$. 
$N_\text{sig}$ for the five parameter sets in the B-model are shown in Table~\ref{Table:Nsignal2l}. 
The observed (expected) upper limits at 95\% confidence level (CL) on $N_\text{sig}$ are 4.4 (5.0) and 3.8 (4.0) for $pp\to e^+ e^-$ and $pp\to \mu^+ \mu^-$, 
which disfavors the parameter sets B$^+$ and B$^\text{L}$ at 95\% CL. 
The parameter sets A1 and A2 are disfavored by the expected upper limits at 95\% CL for $pp\to e^+ e^-$ and $pp\to \mu^+ \mu^-$. 
\begin{table}[htbh]\centering
    \begin{tabular}{cc|ccc|ccc}\hline
        &&\multicolumn{3}{c}{$pp\to e^+ e^-$}&\multicolumn{3}{|c}{$pp\to \mu^+ \mu^-$}\\
        \hline
                    & & & $m_\text{KK}$ & & & $m_\text{KK}$ & \\ 
                    & & 11 TeV & 13 TeV & 15 TeV & 11 TeV & 13 TeV & 15 TeV \\
        \hline
                    & 0.11 &     & 5.7 &     &     & 5.0 &     \\
         $\theta_H$ & 0.10 & 6.1 & 3.9 & 2.7 & 4.9 & 3.1 & 2.1 \\
                    & 0.09 &     & 2.4 &     &     & 1.6 &     \\
        \hline
    \end{tabular}
    \caption{
        The number of signal events in the GHU B-model $N_\text{sig}\equiv N_\text{GHU}-N_\text{SM}$, where $N_\text{SM}=1.2$ for $pp\to e^+ e^-$ and $N_\text{SM}=1.8$ for $pp\to \mu^+ \mu^-$. 
        The second and third columns show $N_\text{sig}$ for $pp\to e^+ e^-$ and $pp\to \mu^+ \mu^-$ processes, respectively. 
        The observed (expected) upper limits at 95\% CL on $N_\text{sig}$ are 4.4 (5.0) and 3.8 (4.0) for $pp\to e^+ e^-$ and $pp\to \mu^+ \mu^-$, respectively. }
    \label{Table:Nsignal2l}
\end{table}

Next, we consider the constraints on the $W'$ bosons from the $pp\to e\nu$ and $\mu\nu$ processes. 
We consider the constraint from the ATLAS group~\cite{ATLAS:2019lsy}. 
The CRs and SRs are not specified there. 
The differential cross sections in the GHU model are larger than that in the SM 
above $M_\text{T}\sim 2300$~GeV for the parameter sets (B$^\text{L}$, B, B$^\text{H}$) and (B$^+$, B, B$^-$). 
Therefore we take the SRs as SR:~[2300,~6000]~GeV for $pp\to e\nu$ and SR:~[2400,~6000]~GeV for $pp\to \mu\nu$, respectively. 
We take the CRs as CR: [310,~1030]~GeV for $pp\to e\nu$ and CR: [300,~1050]~GeV for $pp\to \mu\nu$. 
$N_\text{SM}=6.8$ for $pp\to e\nu$ and $N_\text{SM}=3.8$ for $pp\to \mu\nu$. 
$N_\text{sig}$s for the five parameter sets are shown in Table~\ref{Table:Nsignal1l}. 
The observed (expected) upper limits at 95\% CL on $N_\text{sig}$ are 3.4 (8.4) and 8.6 (7.7) for $pp\to e\nu$ and $pp\to \mu\nu$, respectively. 
Therefore the parameter sets B, B$^+$ and B$^\text{L}$ are disfavored by the expected upper limits at 95\% CL for $pp\to e\nu$. 
The parameter sets (A1, A2, A3) are not excluded from the $pp\to e\nu$ and $pp\to \mu\nu$ processes because of the large $W'$ mass and small gauge couplings. 
\begin{table}[htbh]\centering
    \begin{tabular}{cc|ccc|ccc}\hline
        &&\multicolumn{3}{c}{$pp\to e\nu$}&\multicolumn{3}{|c}{$pp\to \mu\nu$}\\
        \hline
                    & & & $m_\text{KK}$ & & & $m_\text{KK}$ & \\ 
                    & & 11 TeV & 13 TeV & 15 TeV & 11 TeV & 13 TeV & 15 TeV \\
        \hline
                    & 0.11 &    & 17 &     &    & 10  &     \\
         $\theta_H$ & 0.10 & 22 & 12 & 6.9 & 13 & 7.3 & 4.3 \\
                    & 0.09 &    & 6.9 &    &    & 4.2 &     \\
        \hline
    \end{tabular}
    \caption{
        The number of signal events in the GHU B-model $N_\text{sig}\equiv N_\text{GHU}-N_\text{SM}$, where $N_\text{SM}=6.8$ for $pp\to e\nu$ and $N_\text{SM}=3.8$ for $pp\to \mu\nu$. 
        The second and third columns show $N_\text{sig}$ for $pp\to e\nu$ and $pp\to \mu\nu$ processes. 
        The observed (expected) upper limits at 95\% CL on $N_\text{sig}$ are 3.4 (8.4) and 8.6 (7.7) for $pp\to e\nu$ and $pp\to \mu\nu$, respectively. }
    \label{Table:Nsignal1l}
\end{table}

\section{For Future LHC Experiments}
In this section, we estimate the numbers of events of the $pp\to \{W, W'\} \to l\nu$ and $pp\to \{\gamma, Z, Z'\} \to l^+l^-$ processes at $\sqrt{s}=14$~TeV 
with the luminosity 300~fb$^{-1}$ (LHC Run 3) and 3000~fb$^{-1}$ (HL-LHC). 
Other background processes are ignored and the acceptance and efficiency are not taken into account. 
The numbers of events $N_\text{GHU}$ and $N_\text{SM}$ are estimated by integrating the differential cross sections times luminosity from $m_\text{min}$ to 6~TeV, 
where $m_\text{min}$ is determined by the condition 
\begin{align}
    \left.\frac{d\sigma_{pp\to l\nu}^\text{GHU}}{dm_\text{T}}\right|_{m_\text{T}=m_\text{min}}=\left.\frac{d\sigma_{pp\to l\nu}^\text{SM}}{dm_\text{T}}\right|_{m_\text{T}=m_\text{min}},
\end{align}
or
\begin{align}
    \left.\frac{d\sigma_{pp\to l^+l^-}^\text{GHU}}{dm_{ll}}\right|_{m_{ll}=m_\text{min}}=\left.\frac{d\sigma_{pp\to l^+l^-}^\text{SM}}{dm_{ll}}\right|_{m_{ll}=m_\text{min}}.
\end{align}

The discovery significance is given by~\cite{Zyla:2020zbs,Cowan:2010js}
\begin{align}
    Z\equiv\sqrt{2\left(N_\text{GHU}\:\ln\frac{N_\text{GHU}}{N_\text{SM}}+N_\text{SM}-N_\text{GHU}\right)}.
    \label{significance}
\end{align}
Although the formula \eqref{significance} is derived in the large number limit, 
we use this formula independent of the numbers of events for simplicity. 
For a given $Z$, the corresponding $p$-value, which is defined as the probability of obtaining a larger excess, is the same as 
$p=1-F(\mu+Z\sigma)$, where $F$ is the Gaussian cumulative distribution function, $\mu$ is a mean and $\sigma$ is a standard deviation. 
Usually, an excess larger than $5\sigma$ is qualified as a discovery. 
Therefore, we estimate the parameter set which gives $Z\simeq 5$. 
The $p=0.05$ corresponds to the discovery significance $Z=1.64$, and a model is allowed at a 95\% CL for $Z<1.64$~\cite{Cowan:2010js}. 
We calculate the $p$-value by assuming that  events follow the Poisson distribution. 

In the A-model, the differential cross section of the $pp \to l\nu$ process is almost equal to that in the SM for $m_\text{KK}>8$~TeV at $\sqrt{s}=14$~TeV. 
For the parameter set A3, the differential cross section of $pp \to e^+e^-$ process is larger above $m_\text{min}=2.287$~TeV. 
With the luminosity 300~fb$^{-1}$, $N_\text{GHU}=70.1$, $N_\text{SM}=28.8$ and the corresponding discovery significance is 6.49. 
The numbers of events and discovery significance of $pp \to \mu^+\mu^-$ process are smaller than those of the $pp \to e^+e^-$ process. 

\begin{table}[thbh]\centering
    \begin{tabular}{cc|c|ccccc}\hline
        $\theta_H$ & $m_\text{KK}$ (TeV) & $z_L$ & $m_\text{min}$ (TeV) & $N_\text{GHU}$ & $N_\text{SM}$ & significance \\
        \hline
        0.09 & 15 & $3.196\times 10^{12}$ & 2.482 & 63.0 & 30.8 & 5.08 \\
        0.08 & 17 & $5.001\times 10^{12}$ & 2.831 & 22.5 & 12.0 & 2.69 \\
        0.07 & 19 & $1.431\times 10^{12}$ & 3.263 & 6.86 & 3.91 & 1.34 \\
        0.06 & 22 & $9.479\times 10^{11}$ & 3.854 & 1.39 & 0.877& 0.508 \\
        \hline
    \end{tabular}
    \caption{The model parameters, lower limits of integral, numbers of events and discovery significance of the $pp \to e\nu$ process at $\sqrt{s}=14$~TeV with the luminosity 300~fb$^{-1}$. }
    \label{Prediction_Wprime}
    \vspace{1em}
    \begin{tabular}{cc|c|ccccc}\hline
        $\theta_H$ & $m_\text{KK}$ (TeV) & $z_L$ & $m_\text{min}$ (TeV) & $N_\text{GHU}$ & $N_\text{SM}$ & significance \\
        \hline
        0.09 & 15 & $3.196\times 10^{12}$ & 2.790 & 19.0 & 7.96 & 3.33 \\
        0.08 & 17 & $5.001\times 10^{12}$ & 3.142 & 7.29 & 3.40 & 1.83 \\
        0.07 & 19 & $1.431\times 10^{12}$ & 3.568 & 2.43 & 1.26 & 0.925 \\
        0.06 & 22 & $9.479\times 10^{11}$ & 4.152 & 0.562 & 0.335 & 0.357 \\
        \hline
    \end{tabular}
    \caption{The model parameters, lower limits of integral, numbers of events and discovery significance of the $pp \to e^+e^-$ process at $\sqrt{s}=14$~TeV with the luminosity 300~fb$^{-1}$. }
    \label{Prediction_Zprime}
\end{table}

For the B-model, the discovery significance of the $pp \to e\nu$ process is larger than that of the $pp \to \mu\nu$ and $pp \to l^+l^-$ processes for the same parameter sets considered bellow. 
We choose  parameter sets  $z_L\simeq O(10^{12})$ with integral $m_\text{KK}/$TeV. 
The results for the $pp \to e\nu$ and $pp \to e^+e^-$ processes are summarized in Tables~\ref{Prediction_Wprime} and \ref{Prediction_Zprime}. 
The masses and decay widths of the $W'$ and $Z'$ for those parameters are shown in Tables~\ref{Table:Mass-Width-Vector-Bosons-2},~\ref{Table:Couplings_mKK=15},~\ref{Table:Couplings_mKK=17},~\ref{Table:Couplings_mKK=19},~\ref{Table:Couplings_mKK=22}.

\begin{table}[thb]
    \centering
    \begin{tabular}{cc|ccccccc|c}
    \hline
    $\theta_H$&$m_\text{KK}$&$m_{W^{(1)}}$&$m_{W_R^{(1)}}$&$\Gamma_{W^{(1)}}$&$\Gamma_{W_R^{(1)}}$&$\Gamma_{Z^{(1)}}$&$\Gamma_{Z_R^{(1)}}$&$\Gamma_{\gamma^{(1)}}$&Table\\
    \mbox{[rad]}&[TeV]&[TeV]&[TeV]&[TeV]&[TeV]&[TeV]&[TeV]&[TeV]&\\ 
    \hline 
    0.09&15&11.74&11.48&12.17&0.458&9.712&0.989&4.007&\ref{Table:Couplings_mKK=15}\\
    0.08&17&13.31&13.01&13.98&0.527&11.16&1.132&4.603&\ref{Table:Couplings_mKK=17}\\
    0.07&19&14.89&14.54&14.98&0.563&11.96&1.231&4.952&\ref{Table:Couplings_mKK=19}\\
    0.06&22&17.24&16.84&17.10&0.642&13.66&1.413&5.662&\ref{Table:Couplings_mKK=22}\\
    \hline
    \end{tabular}
     \caption{
     Masses and decay widths of $W^{(1)}$, $W_R^{(1)}$, $Z^{(1)}$, $Z_R^{(1)}$ and $\gamma^{(1)}$ are listed. 
     $m_{W^{(1)}} \simeq m_{Z^{(1)}} \simeq m_{\gamma^{(1)}}$ with 4 digits of precision. 
     $m_{W_R^{(1)}}$ and $m_{Z_R^{(1)}}$ are exactly equal. 
     }
    \label{Table:Mass-Width-Vector-Bosons-2}
\end{table}

\begin{table}[thb]
    \centering
    \begin{tabular}{ccccccc}
        \hline
        $ff'$ &$g_{Wff'}^L$ & $g_{Wff'}^R$ & $g_{W^{(1)}ff'}^L$ & $g_{W^{(1)}ff'}^R$ & $g_{W_R^{(1)}ff'}^L$ & $g_{W_R^{(1)}ff'}^R$\\
        \hline
        $e \nu_e$     & 0.9981 & 0 & 5.9277 & 0 & 0.0121 & 0 \\
        $\mu \nu_\mu$ & 0.9981 & 0 & 5.6554 & 0 & 0.0116 & 0 \\
        $\tau \nu_\tau$&0.9981 & 0 & 5.4764 & 0 & 0.0113 & 0 \\
        $ud$ & 0.9981 & 0 & 5.7462 & 0 & 0.0118 & 0 \\
        $cs$ & 0.9981 & 0 & 5.5457 & 0 & 0.0114 & 0 \\
        $tb$ & 0.9983 & 0 & 4.6551 & 0 & 0.0097 & $-$0.0330 \\
        \hline
    \end{tabular}\\ \vspace{1em}
    \begin{tabular}{ccccccccc}
        \hline
        $f$ & $g_{Zf}^L$ & $g_{Zf}^R$ & $g_{Z^{(1)}f}^L$ & $g_{Z^{(1)}f}^R$ & $g_{Z_R^{(1)}f}^L$ & $g_{Z_R^{(1)}f}^R$ & $g_{\gamma^{(1)}f}^L$ & $g_{\gamma^{(1)}f}^R$\\
        \hline
        $\nu_e$    &\ \ 0.5690 & 0 &\ \ 3.3812 & 0 & $-$1.0668 & 0 & 0 & 0 \\
        $\nu_\mu$  &\ \ 0.5690 & 0 &\ \ 3.2259 & 0 & $-$1.0200 & 0 & 0 & 0 \\
        $\nu_\tau$ &\ \ 0.5690 & 0 &\ \ 3.1238 & 0 & $-$0.9892 & 0 & 0 & 0 \\[0.5em]
        $e$    &$-$0.3059 &\ \ 0.2630 & $-$1.8180 & $-$0.0562 & $-$1.0770 & 0 & $-$2.8470 &\ \ 0.1030 \\
        $\mu$  &$-$0.3059 &\ \ 0.2630 & $-$1.7345 & $-$0.0562 & $-$1.0297 & 0 & $-$2.7160 &\ \ 0.1030 \\
        $\tau$ &$-$0.3059 &\ \ 0.2630 & $-$1.6796 & $-$0.0562 & $-$0.9986 & 0 & $-$2.6303 &\ \ 0.1029 \\[0.5em]
        $u$ &\ \ 0.3936 & $-$0.1754 &\ \ 2.2675 &\ \ 0.0375 &\ \ 0.3518 & 0 &\ \ 1.8399 & $-$0.0687 \\
        $c$ &\ \ 0.3936 & $-$0.1754 &\ \ 2.1188 &\ \ 0.0375 &\ \ 0.3401 & 0 &\ \ 1.7757 & $-$0.0687 \\
        $t$ &\ \ 0.3939 & $-$0.1751 &\ \ 1.8369 & $-$0.3128 &\ \ 0.2878 & $-$0.7078 &\ \ 1.4907 &\ \ 0.5749 \\[0.5em]
        $d$ & $-$0.4813 &\ \ 0.0877 & $-$2.7726 &\ \ 0.1133 &\ \ 0.3419 & $-$0.1758 & $-$0.9200 & $-$0.2062 \\
        $s$ & $-$0.4813 &\ \ 0.0877 & $-$2.6759 &\ \ 0.1428 &\ \ 0.3305 & $-$0.2150 & $-$0.8878 & $-$0.2599 \\
        $b$ & $-$0.4813 &\ \ 0.0877 & $-$2.2461 &\ \ 0.2844 &\ \ 0.2798 & $-$0.4024 & $-$0.7452 & $-$0.5179 \\
        \hline
    \end{tabular}
    \caption{
        Coupling constants of vector bosons are listed for $\theta_H=0.09$ and $m_\text{KK}=15$~TeV in Table~\ref{Table:Mass-Width-Vector-Bosons-2},
        where $\sin^2\theta_W^0=0.2307$. 
        The above table shows the couplings of $W'$ bosons to fermions in units of $g_w/\sqrt{2}$. 
        The below table shows the couplings of $Z'$ bosons to fermions in units of $g_w=e/\sin\theta_W^0$. 
        Their corresponding $\gamma$ boson coupling constants are the same as those in the SM. The values less than $10^{-4}$ are written as $0$.
        }
    \label{Table:Couplings_mKK=15}
\end{table}

\begin{table}[thb]
    \centering
    \begin{tabular}{ccccccc}
        \hline
        $ff'$ &$g_{Wff'}^L$ & $g_{Wff'}^R$ & $g_{W^{(1)}ff'}^L$ & $g_{W^{(1)}ff'}^R$ & $g_{W_R^{(1)}ff'}^L$ & $g_{W_R^{(1)}ff'}^R$\\
        \hline
        $e \nu_e$     & 0.9985 & 0 & 5.9655 & 0 & 0.0097 & 0 \\
        $\mu \nu_\mu$ & 0.9985 & 0 & 5.6937 & 0 & 0.0092 & 0 \\
        $\tau \nu_\tau$&0.9985 & 0 & 5.5155 & 0 & 0.0090 & 0 \\
        $ud$ & 0.9985 & 0 & 5.7842 & 0 & 0.0094 & 0 \\
        $cs$ & 0.9985 & 0 & 5.5844 & 0 & 0.0091 & 0 \\
        $tb$ & 0.9987 & 0 & 4.7041 & 0 & 0.0077 & $-$0.0327 \\
        \hline
    \end{tabular}\\ \vspace{1em}
    \begin{tabular}{ccccccccc}
        \hline
        $f$ & $g_{Zf}^L$ & $g_{Zf}^R$ & $g_{Z^{(1)}f}^L$ & $g_{Z^{(1)}f}^R$ & $g_{Z_R^{(1)}f}^L$ & $g_{Z_R^{(1)}f}^R$ & $g_{\gamma^{(1)}f}^L$ & $g_{\gamma^{(1)}f}^R$\\
        \hline
        $\nu_e$    &\ \ 0.5693 & 0 &\ \ 3.4026 & 0 & $-$1.0757 & 0 & 0 & 0 \\
        $\nu_\mu$  &\ \ 0.5693 & 0 &\ \ 3.2475 & 0 & $-$1.0288 & 0 & 0 & 0 \\
        $\nu_\tau$ &\ \ 0.5693 & 0 &\ \ 3.1459 & 0 & $-$0.9981 & 0 & 0 & 0 \\[0.5em]
        $e$    &$-$0.3061 &\ \ 0.2632 & $-$1.8294 & $-$0.0559 & $-$1.0837 & 0 & $-$2.8658 &\ \ 0.1022 \\
        $\mu$  &$-$0.3061 &\ \ 0.2632 & $-$1.7461 & $-$0.0559 & $-$1.0365 & 0 & $-$2.7352 &\ \ 0.1022 \\
        $\tau$ &$-$0.3061 &\ \ 0.2632 & $-$1.6914 & $-$0.0559 & $-$1.0056 & 0 & $-$2.6496 &\ \ 0.1021 \\[0.5em]
        $u$ &\ \ 0.3938 & $-$0.1754 &\ \ 2.2823 &\ \ 0.0372 &\ \ 0.3534 & 0 &\ \ 1.8525 & $-$0.0681 \\
        $c$ &\ \ 0.3938 & $-$0.1754 &\ \ 2.2035 &\ \ 0.0372 &\ \ 0.3417 & 0 &\ \ 1.7885 & $-$0.0681 \\
        $t$ &\ \ 0.3940 & $-$0.1752 &\ \ 1.8561 & $-$0.3108 &\ \ 0.2902 & $-$0.7020 &\ \ 1.5067 &\ \ 0.5702 \\[0.5em]
        $d$ & $-$0.4815 &\ \ 0.0877 & $-$2.7908 &\ \ 0.1127 &\ \ 0.3455 & $-$0.1751 & $-$0.9262 & $-$0.2052 \\
        $s$ & $-$0.4815 &\ \ 0.0877 & $-$2.6944 &\ \ 0.1421 &\ \ 0.3341 & $-$0.2142 & $-$0.8943 & $-$0.2588 \\
        $b$ & $-$0.4815 &\ \ 0.0877 & $-$2.2696 &\ \ 0.2833 &\ \ 0.2837 & $-$0.4014 & $-$0.7533 & $-$0.5161 \\
        \hline
    \end{tabular}
    \caption{
        Coupling constants of vector bosons are listed for $\theta_H=0.08$ and $m_\text{KK}=17$~TeV in Table~\ref{Table:Mass-Width-Vector-Bosons-2},
        where $\sin^2\theta_W^0=0.2308$. 
        }
    \label{Table:Couplings_mKK=17}
\end{table}

\begin{table}[thb]
    \centering
    \begin{tabular}{ccccccc}
        \hline
        $ff'$ &$g_{Wff'}^L$ & $g_{Wff'}^R$ & $g_{W^{(1)}ff'}^L$ & $g_{W^{(1)}ff'}^R$ & $g_{W_R^{(1)}ff'}^L$ & $g_{W_R^{(1)}ff'}^R$\\
        \hline
        $e \nu_e$     & 0.9988 & 0 & 5.8583 & 0 & 0.0073 & 0 \\
        $\mu \nu_\mu$ & 0.9988 & 0 & 5.5850 & 0 & 0.0069 & 0 \\
        $\tau \nu_\tau$&0.9988 & 0 & 5.4046 & 0 & 0.0067 & 0 \\
        $ud$ & 0.9987 & 0 & 5.6763 & 0 & 0.0070 & 0 \\
        $cs$ & 0.9987 & 0 & 5.4745 & 0 & 0.0068 & 0 \\
        $tb$ & 0.9989 & 0 & 4.5622 & 0 & 0.0057 & $-$0.0336 \\
        \hline
    \end{tabular}\\ \vspace{1em}
    \begin{tabular}{ccccccccc}
        \hline
        $f$ & $g_{Zf}^L$ & $g_{Zf}^R$ & $g_{Z^{(1)}f}^L$ & $g_{Z^{(1)}f}^R$ & $g_{Z_R^{(1)}f}^L$ & $g_{Z_R^{(1)}f}^R$ & $g_{\gamma^{(1)}f}^L$ & $g_{\gamma^{(1)}f}^R$\\
        \hline
        $\nu_e$    &\ \ 0.5695 & 0 &\ \ 3.3413 & 0 & $-$1.0646 & 0 & 0 & 0 \\
        $\nu_\mu$  &\ \ 0.5695 & 0 &\ \ 3.1854 & 0 & $-$1.0173 & 0 & 0 & 0 \\
        $\nu_\tau$ &\ \ 0.5695 & 0 &\ \ 3.0825 & 0 & $-$0.9859 & 0 & 0 & 0 \\[0.5em]
        $e$    &$-$0.3062 &\ \ 0.2633 & $-$1.7964 & $-$0.0572 & $-$1.0586 & 0 & $-$2.8150 &\ \ 0.1045 \\
        $\mu$  &$-$0.3062 &\ \ 0.2633 & $-$1.7126 & $-$0.0572 & $-$1.0114 & 0 & $-$2.6836 &\ \ 0.1045 \\
        $\tau$ &$-$0.3062 &\ \ 0.2633 & $-$1.6573 & $-$0.0571 & $-$0.9803 & 0 & $-$2.5969 &\ \ 0.1044 \\[0.5em]
        $u$ &\ \ 0.3939 & $-$0.1755 &\ \ 2.2396 &\ \ 0.0381 &\ \ 0.3463 & 0 &\ \ 1.8184 & $-$0.0697 \\
        $c$ &\ \ 0.3939 & $-$0.1755 &\ \ 2.1600 &\ \ 0.0381 &\ \ 0.3346 & 0 &\ \ 1.7537 & $-$0.0697 \\
        $t$ &\ \ 0.3941 & $-$0.1754 &\ \ 1.8000 & $-$0.3199 &\ \ 0.2814 & $-$0.7201 &\ \ 1.4615 &\ \ 0.5861 \\[0.5em]
        $d$ & $-$0.4817 &\ \ 0.0878 & $-$2.7385 &\ \ 0.1144 &\ \ 0.3404 & $-$0.1779 & $-$0.9092 & $-$0.2083 \\
        $s$ & $-$0.4817 &\ \ 0.0878 & $-$2.6412 &\ \ 0.1440 &\ \ 0.3289 & $-$0.2173 & $-$0.8769 & $-$0.2622 \\
        $b$ & $-$0.4817 &\ \ 0.0878 & $-$2.2010 &\ \ 0.2867 &\ \ 0.2766 & $-$0.4063 & $-$0.7307 & $-$0.5223 \\
        \hline
    \end{tabular}
    \caption{
         Coupling constants of vector bosons are listed for $\theta_H=0.07$ and $m_\text{KK}=19$~TeV in Table~\ref{Table:Mass-Width-Vector-Bosons-2},
         where $\sin^2\theta_W^0=0.2309$. 
         }
    \label{Table:Couplings_mKK=19}
\end{table}

\begin{table}[thb]
    \centering
    \begin{tabular}{ccccccc}
        \hline
        $ff'$ &$g_{Wff'}^L$ & $g_{Wff'}^R$ & $g_{W^{(1)}ff'}^L$ & $g_{W^{(1)}ff'}^R$ & $g_{W_R^{(1)}ff'}^L$ & $g_{W_R^{(1)}ff'}^R$\\
        \hline
        $e \nu_e$     & 0.9992 & 0 & 5.8233 & 0 & 0.0053 & 0 \\
        $\mu \nu_\mu$ & 0.9992 & 0 & 5.5496 & 0 & 0.0051 & 0 \\
        $\tau \nu_\tau$&0.9992 & 0 & 5.3685 & 0 & 0.0049 & 0 \\
        $ud$ & 0.9992 & 0 & 5.6412 & 0 & 0.0051 & 0 \\
        $cs$ & 0.9992 & 0 & 5.4387 & 0 & 0.0050 & 0 \\
        $tb$ & 0.9993 & 0 & 4.5149 & 0 & 0.0042 & $-$0.0339 \\
        \hline
    \end{tabular}\\ \vspace{1em}
    \begin{tabular}{ccccccccc}
        \hline
        $f$ & $g_{Zf}^L$ & $g_{Zf}^R$ & $g_{Z^{(1)}f}^L$ & $g_{Z^{(1)}f}^R$ & $g_{Z_R^{(1)}f}^L$ & $g_{Z_R^{(1)}f}^R$ & $g_{\gamma^{(1)}f}^L$ & $g_{\gamma^{(1)}f}^R$\\
        \hline
        $\nu_e$    &\ \ 0.5697 & 0 &\ \ 3.3212 & 0 & $-$1.0541 & 0 & 0 & 0 \\
        $\nu_\mu$  &\ \ 0.5697 & 0 &\ \ 3.1651 & 0 & $-$1.0069 & 0 & 0 & 0 \\
        $\nu_\tau$ &\ \ 0.5697 & 0 &\ \ 3.0618 & 0 & $-$0.9756 & 0 & 0 & 0 \\[0.5em]
        $e$    &$-$0.3063 &\ \ 0.2634 & $-$1.7854 & $-$0.0577 & $-$1.0585 & 0 & $-$2.7988 &\ \ 0.1054 \\
        $\mu$  &$-$0.3063 &\ \ 0.2634 & $-$1.7015 & $-$0.0577 & $-$1.0111 & 0 & $-$2.6672 &\ \ 0.1054 \\
        $\tau$ &$-$0.3063 &\ \ 0.2634 & $-$1.6456 & $-$0.0576 & $-$0.9797 & 0 & $-$2.5912 &\ \ 0.1053 \\[0.5em]
        $u$ &\ \ 0.3941 & $-$0.1756 &\ \ 2.2255 &\ \ 0.0385 &\ \ 0.3438 & 0 &\ \ 1.8075 & $-$0.0702 \\
        $c$ &\ \ 0.3941 & $-$0.1756 &\ \ 2.1456 &\ \ 0.0385 &\ \ 0.3320 & 0 &\ \ 1.7426 & $-$0.0702 \\
        $t$ &\ \ 0.3942 & $-$0.1755 &\ \ 1.7812 & $-$0.3235 &\ \ 0.2782 & $-$0.7264 &\ \ 1.4467 &\ \ 0.5919 \\[0.5em]
        $d$ & $-$0.4819 &\ \ 0.0878 & $-$2.7214 &\ \ 0.1149 &\ \ 0.3395 & $-$0.1789 & $-$0.9037 & $-$0.2093 \\
        $s$ & $-$0.4819 &\ \ 0.0878 & $-$2.6238 &\ \ 0.1446 &\ \ 0.3279 & $-$0.2184 & $-$0.8713 & $-$0.2634 \\
        $b$ & $-$0.4819 &\ \ 0.0878 & $-$2.1781 &\ \ 0.2879 &\ \ 0.2748 & $-$0.4083 & $-$0.7233 & $-$0.5245 \\
        \hline
    \end{tabular}
    \caption{
        Coupling constants of vector bosons are listed for $\theta_H=0.06$ and $m_\text{KK}=22$~TeV in Table~\ref{Table:Mass-Width-Vector-Bosons-2},
        where $\sin^2\theta_W^0=0.2311$. 
        }
    \label{Table:Couplings_mKK=22}
\end{table}

For $\theta_H=0.09$ and $m_\text{KK}=15$~TeV, 
it is found that $m_\text{min}=2.482$~TeV, $N_\text{GHU}=63$ and $N_\text{SM}=31$ at the LHC Run 3, where the discovery significance is $Z=5.08$. 
When about 63 events are observed for the $pp \to e\nu$ process in the transverse mass range $2.5\text{~TeV} \lesssim m_\text{T}\lesssim 6.0$~TeV at the LHC Run 3, 
the discovery of new physics is expected, and the GHU model becomes viable. 
The numbers of events of the GHU (SM) in each bin are 67.5 (92.5), 29.7 (21.6), 24.1 (7.1), 6.3 (0.56) and 1.4 (0.05)
for [2000, 2500] GeV, [2500, 3000] GeV, [3000, 4000] GeV, [4000, 5000] GeV and [5000, 6000] GeV, respectively. 

By interpolating the numerical results shown in Table~\ref{Prediction_Wprime}, we obtain $Z=1.64$ for $\theta_H=0.0727$ and $m_\text{KK}=18.44$~TeV, where $N_\text{GHU}=9.75$ and $N_\text{SM}=5.48$. 
Hence, as a rough estimate, the upper limit of the KK scale testable at the LHC Run 3 is $m_\text{KK}\simeq18.44$~TeV. 
At the HL-LHC, the total integrated luminosity 3000~fb$^{-1}$ of data is going to be collected~\cite{Azzi:2019yne,Cepeda:2019klc,CidVidal:2018eel}. 
For $\theta_H=0.06$ and $m_\text{KK}=22$~TeV, 
$N_\text{GHU}=13.9$ and $N_\text{SM}=8.77$ at $\sqrt{s}=14$~TeV with the 3000 fb$^{-1}$ luminosity. 
The discovery significance is $Z=1.61$ and the corresponding $p$-value is $p=0.0627$. 
The GHU B-model is testable up to $m_\text{KK}\simeq22$~TeV. 

We add that backgrounds coming from other processes, acceptance and efficiency have not been taken into account in the evaluation in this section.

\section{Summary and Discussions}
In this paper we studied the $pp\to \{W, W'\} \to l\nu$ and $pp\to \{\gamma, Z, Z'\} \to l^+l^-$ ($l=e,\mu$) processes in the $SU(3)_C\times SO(5)\times U(1)_X$ GHU models. 
Due to the behavior of the wave functions of various fields in the fifth dimension, the $Z'$ couplings of right- and left-handed quarks become relatively large in the GHU A-model and B-model, respectively. 
The largest decay width of the $Z'$ bosons is found to be $\Gamma_{\gamma^{(1)}}/m_{\gamma^{(1)}}\sim 0.1$ in the A-model and $\Gamma_{Z^{(1)}}/m_{Z^{(1)}}\sim 0.8$ in the B-model. 
The $W'$ couplings of left-handed quarks also become large in the GHU B-model, and $\Gamma_{W^{(1)}}/m_{W^{(1)}}\sim 1$. 
In contrast to it, the $W^{(1)}$ couplings in the A-model remain small with $\Gamma_{W^{(1)}}/m_{W^{(1)}}\sim 0.04$. 

The differential cross sections of the $pp\to l^+l^-$ processes in the GHU models are smaller than those in the SM for the invariant mass $m_{ll}\lesssim 2$~TeV. 
From the searches for events in the dilepton final states at $\sqrt{s}=13$~TeV with up to 140 fb$^{-1}$ of data~\cite{ATLAS:2020yat}, 
the A-model is constrained as $\theta_H\lesssim0.08$ and $m_\text{KK}\gtrsim9.5$~TeV, and 
the B-model is constrained as $\theta_H\lesssim0.10$ and $m_\text{KK}\gtrsim13$~TeV. 

The differential cross sections of the $pp\to l\nu$ processes in the GHU B-model are also smaller than those in the SM for the transverse mass $m_\text{T}\lesssim 2$~TeV. 
The constraint on the B-model from the searches for events in the lepton and missing transverse mass final states at $\sqrt{s}=13$~TeV with up to 140 fb$^{-1}$ of data~\cite{ATLAS:2019lsy} 
is severe compared with the constraint from those in the dilepton final states. 
The constraint on the B-model is $\theta_H<0.10$ and $m_\text{KK}>13$~TeV. 
The A-model is consistent with the experimental data due to the narrow decay width of the $W^{(1)}$ boson. 

At $\sqrt{s}=14$~TeV with the luminosity 300~fb$^{-1}$, 
signals of $Z'$ bosons in the A-model can be seen in the $pp \to l^+ l^-$ processes for $\theta_H=0.08$ and $m_\text{KK}=9.5$~TeV. 
In the B-model, signals of $W'$ bosons can be seen in the $pp \to e\nu$ process for $m_\text{KK}\lesssim 15$~TeV and $\theta_H\gtrsim0.09$. 
The upper limit of the KK scale in the B-model can be pushed to $m_\text{KK}\simeq 18$~TeV with the luminosity 300~fb$^{-1}$ and to $m_\text{KK}\simeq 22$~TeV with luminosity 3000~fb$^{-1}$. 

At the ILC, the effects of $Z'$ bosons in fermion pair production processes can be seen even at $\sqrt{s}=250$~GeV by using polarized electron and positron beams. 
For $m_\text{KK}< 22$~TeV, the deviations of the cross section for the $e^+ e^- \to \mu^+\mu^-$ process in the GHU B-model from that in the SM are $O(1)$\% for a left-handed electron beam and $O(0.1)$\% with a right-handed electron beam at $\sqrt{s}=250$~GeV, 
where the statistical uncertainty with the 250~fb$^{-1}$ luminosity data is about 0.1\%~\cite{Funatsu:2020haj}. 
To reduce theoretical uncertainties, further studies beyond the tree-level are necessary. 

Collider physics of radions, KK gravitons and KK gluons are also important subjects in models defined on a higher dimensional spacetime~\cite{Agashe:2020wph,Escribano:2021jne}.
For instance,  KK gluons mediate dijet and $t\bar{t}$ production processes at hadron colliders. 
In the $t\bar{t}$ production processes, the forward-backward asymmetry~\cite{CDF:2017cvy,CMS:2019kzp} and the charge asymmetry~\cite{ATLAS:2017gkv,ATLAS:2019czt} have been measured, which so far have been consistent with the SM predictions. 
Effects of KK gluons in the $pp\to t\bar{t}$ process need be studied in the GHU models as well. 
KK gluons in the GHU B-model have large couplings to left-handed fermions with a large decay width 
just as  KK photons. 
Broad excesses of the differential cross sections are foreseen at the LHC in the process mediated by KK gauge bosons, 
and the polarization dependence of the cross sections should be confirmed at the ILC by using  polarized beams. 
Observing these characteristic signals may provide a strong indication for the existence of the extra dimension.

\section*{Acknowledgments}

This work was supported in part by European Regional Development Fund-Project Engineering Applications of Microworld Physics (Grant No.\ CZ.02.1.01/0.0/0.0/16\_019/0000766) (Y.O.), 
by the National Natural Science Foundation of China (Grants No.\ 11775092, No.\ 11675061, No.\ 11521064, No.\ 11435003 and No.\ 11947213) (S.F.), by the International Postdoctoral Exchange Fellowship Program (S.F.), 
and by Japan Society for the Promotion of Science, Grants-in-Aid for Scientific Research, Grant No.\ JP19K03873 (Y.H.) and Grant No.\ JP18H05543 (N.Y.).

\end{document}